\documentclass{aa}  

\usepackage{graphicx}
\usepackage{txfonts}
\usepackage{color}                       % For color text: \color command (adicionado)
\usepackage{url}                         % For breaking URLs easily trough lines (adicionado)
\usepackage{rotating}                    % (adicionado)   
\usepackage{lscape}                      % (adicionado)
\begin{document} 

   \title{Solar type II radio bursts associated with CME expansions as shown by EUV waves}

   %\subtitle{I. Overviewing the $\kappa$-mechanism}

   \author{R. D. Cunha-Silva
%           \inst{1}
          \and
          F. C. R. Fernandes%\inst{1}%\fnmsep\thanks{Just to show the usage
          %of the elements in the author field}
          \and
          C. L. Selhorst%\inst{1}
          }
     
    \titlerunning{Solar type II radio bursts associated with CME expansions as shown by EUV waves}
    \authorrunning{Cunha-Silva et al.}
     
   \institute{IP\&D - Universidade do Vale do Para\'iba - UNIVAP, Av. Shishima Hifumi, 
              2911 - Urbanova, S\~ao Jos\'e dos Campos, SP, Brazil\\ 
              \email{rfldoug@gmail.com}
%          \and
%              IP\&D - Universidade do Vale do Para\'iba - UNIVAP, Av. Shishima Hifumi, 2911 - Urbanova, S\~ao Jos\'e dos Campos, SP 12244-000, Brazil \\
%              \email{c.ptolemy@hipparch.uheaven.space}
%              \thanks{The university of heaven temporarily does not
%                      accept e-mails}
             }

   \date{Received November 21, 2014; accepted April 9, 2015}

% \abstract{}{}{}{}{} 
% 5 {} token are mandatory
 
  \abstract
  %{} leave it empty if necessary  
     {}
   {We investigate the physical conditions of the sources of two metric type II bursts associated 
   with coronal mass ejection (CME) expansions with the aim of verifying the relationship between the shocks and the CMEs, 
   by comparing the heights of the radio sources and of the extreme-ultraviolet (EUV) waves associated with the CMEs.}
   {The heights of the EUV waves associated with the events were determined in relation to the 
   wave fronts. The heights of the shocks were estimated by applying two different density models 
   to the frequencies of the type II emissions and compared with the heights of the EUV waves. For the event on 13 June 2010, that 
   included band-splitting, the shock speed was estimated from the frequency drifts of the upper and 
   lower frequency branches of the harmonic lane, taking into account the H/F frequency ratio $f_{\rm{H}}/f_{\rm{F}}=2$. 
   Exponential fits on the intensity maxima of the frequency branches were more consistent with 
   the morphology of the spectrum of this event. For the event on 6 June 2012, that did not include band-splitting and showed 
   a clear fundamental lane on the spectrum, the shock speed was directly estimated from the frequency drift of the fundamental 
   emission, determined by linear fit on the intensity maxima of the lane. For each event, the most appropriate density model was 
   adopted to estimate the physical parameters of the radio source.}
   {The event on 13 June 2010 had a shock speed of 590\,--\,810 km s$^{-1}$, consistent with the average speed of the EUV wave 
   fronts of 610 km s$^{-1}$. The event on 6 June 2012 had a shock speed of 250\,--\,550 km s$^{-1}$, also consistent with the 
   average speed of the EUV wave fronts of 420 km s$^{-1}$. For both events, the heights of the EUV wave revealed to be compatible 
   with the heights of the radio source, assuming a radial propagation of the type-II-emitting shock segment.}
   {}


   \maketitle
   
%
%________________________________________________________________

\section{Introduction}
   
   Radio emission stripes that slowly drift from high to low frequencies in the solar dynamic spectra are known as type II 
   bursts and are generated from plasma oscillations that are attributed to a fast-mode magnetohydrodynamic (MHD) shock 
   waves \citep{Nelson:1985}. The shock propagates outwards through the corona at speeds of 200 to 2000 km s$^{-1}$, as 
   shown by the drifts of the emissions towards lower frequencies owing to the decreasing ambient density. Radial speeds 
   can be deduced from the drift rates ($\sim$0.1\,--\,1.5 MHz s$^{-1}$) of the emissions by using coronal density models.
   
   Type II bursts were suggested to be signatures of shock waves by \citet{Uchida:1960} and \citet{Wild:1962}, and the 
   first evidence of this association came from the interpretation of the Moreton waves \citep{Moreton:1960}, which are 
   large-scale wave-like disturbances in the chromosphere observed in H$\alpha$. The Moreton waves propagate out of the 
   flare site at speeds in the order of those of the shock waves that, at greater heights, cause type II bursts.  

   The emissions are observed at multiples of the electron plasma frequency, given by 
   $f_{\rm{p}}=8.98 \times 10^{-3} \sqrt{n_{\rm{e}}} \ \ [\mathrm{MHz}] \ $, chiefly at the fundamental and second harmonic 
   emissions. Sometimes the third harmonic emission can be observed in the dynamic spectra of type II bursts 
   \citep[see, e.g.,][]{Zlotnik:1998}. 

   In the meter wavelength range, type II bursts typically show a starting frequency for the fundamental emission 
   around 100 MHz. Nevertheless, it is common to observe events with considerably higher starting frequencies 
   \citep[see, e.g.,][]{Vrsnak:2002,Reiner:2003,Vrsnak:2008}. The events last for several minutes ($\sim$1\,--\,15 min), 
   and sometimes the emission stripes are split into two parallel lanes (band-splitting effect); the mechanism that causes 
   this is not fully understood as yet \citep[see, e.g.,][]{Tidman:1966,Kruger:1979,Treumann:1992}. One feasible and 
   potential interpretation of this effect was proposed by \citet{Smerd:1974} in terms of the emission from the upstream 
   and downstream shock regions, which associates this effect with an abrupt density variation at the shock front 
   \citep[see, e.g.,][]{Vrsnak:2001,Vrsnak:2002,Cho:2007,Zimovets:2012}. 

   Although the association between type II bursts and coronal shock waves is well established, the physical relationship among 
   metric type II bursts, flares, and coronal mass ejections (CMEs) is only poorly understood. A causal relationship between metric 
   type II bursts and CMEs is a controversial question \citep[see, e.g.,][]{Cliver:2004,Vrsnak:2008,Prakash:2010}, whereas for type 
   II bursts at dekameter and longer wavelengths there is a consensus that they are driven by CMEs \citep[see, e.g.,][]{Cane:1987,Gopalswamy:2000}. 
   According to \citet{Vrsnak:2008}, the source of the coronal wave seems clear in some events with starting frequencies well below 
   100 MHz, where the CME is accompanied only by a very weak or gradual flare-like energy release. On the other hand, according to these 
   authors, for Moreton-wave-associated type II bursts, that have considerably higher frequencies ($>$ 300 MHz for the harmonic emission), 
   both the CME and the flare are observed. In these cases, the CME and the flare are often tightly related and, particularly during 
   the impulsive phase, motions of the flare plasma take place together with the CME motions, and both phenomena are potential 
   sources of the shocks \citep[for a discussion see, e.g.,][]{Vrsnak:2008,Magdalenic:2012}.
   
   The study of the relationship between type II bursts and CMEs can be improved by observations of propagating brightness fronts in the 
   extreme ultraviolet (EUV), so-called EUV waves \citep[see, e.g.,][]{Moses:1997,Thompson:1998}. The onsets of expanding dimmings that 
   occasionally accompany the EUV waves indicate the source regions of the CME and are related to its fast acceleration phase 
   \citep[see, e.g.,][]{Cliver:2004}. The wave nature of the EUV waves, however, is still under debate. In wave models, they are interpreted 
   as a fast-mode wave, most likely triggered by a CME, and represent the coronal counterpart of the Moreton waves 
   \citep[see, e.g.,][]{Vrsnak:2008,Patsourakos:2012}. In the non-wave models, they are explained by other processes related to the large-scale 
   magnetic field reconfiguration and can be interpreted as a disk projection of the expanding envelope of the CME 
   \citep[see, e.g.,][]{Delannee:1999,Chen:2005,Attrill:2007}, being usually much slower and more diffuse 
   \citep[see, e.g.,][]{Klassen:2000,Warmuth:2004a,Warmuth:2004b}. According to \citet{Klassen:2000}, 90\% of the metric type II bursts are 
   associated with EUV waves, but there is no correlation between their speeds. 

   Given that metric type II bursts are somewhat rarer than the number of flares and CMEs, case studies are helpful in investigating the 
   dynamic behavior of the shocks associated with these phenomena \citep[for a detailed analysis of CME-type II relationship see, e.g.,][]{Magdalenic:2010}. 
   In studying the physical conditions of the sources of two metric type II bursts associated with CME expansions as shown by EUV waves, we aim to 
   verify the relationship between the radio sources and the CMEs mainly in terms of the heights and speeds of the EUV wave fronts and the type II sources. 
   
%__________________________________________________________________

\section{Observations and analysis}

The type II events investigated in this article were observed by two spectrometers from e-CALLISTO (extended-Compound Astronomical Low-cost Low-frequency 
Instrument for Spectroscopy and Transportable Observatories). The dynamic spectra of the type II bursts were digitally recorded with time and frequency 
resolutions of 1.25 ms and 62.5 kHz \citep[see, e.g.,][]{Benz:2009}. The e-CALLISTO data are available at 
\texttt{http://soleil.i4ds.ch/solarradio/data/2002-20yy\_ Callisto/}.  

\subsection{Event on 13 June 2010} 

The event on 13 June 2010 (hereafter event 1) was observed at $\sim$05:37:10 UT by CALLISTO-OOTY (Ooty, India) in the operational 
frequency range of 45\,--\,442 MHz and was included in the Burst Catalog, built by C. Monstein 
(\texttt{http://e-callisto.org/papers/BurstCatalog.pdf}). This event was also observed by the radio spectrograph of Learmonth 
Observatory \citep{Kozarev:2011}, the radio spectrograph of San Vito Observatory \citep{Ma:2011}, the Hiraiso Radio 
Spectrograph \citep{Gopalswamy:2012}, and the radio spectrograph ARTEMIS IV \citep{Kouloumvakos:2014}.  

\begin{figure}[!h]
 \centering
 \includegraphics[width=9cm]{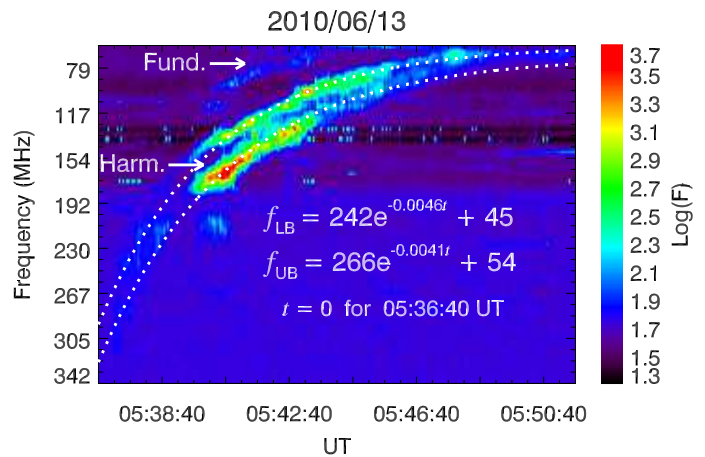}
 \caption{Dynamic spectrum of the type II burst (left panel) observed by CALLISTO-OOTY on June 13, 2010 at $\sim$05:37:10 UT, 
 with exponential fits for the upper (UB) and the lower (LB) frequency branches of the harmonic lane.}
 \label{f1}
\end{figure}

Figure \ref{f1} shows that both the fundamental and harmonic emissions are present in the dynamic spectrum of event 1, with 
the fundamental lane being partially reabsorbed. The bright branches of the harmonic 
emission showed a clear-cut band-splitting effect, with an onset at $\sim$325 MHz. Although partially reabsorbed, the 
band-splitting effect for the fundamental emission was also observed in the spectrum. 

Event 1 was accompanied by a slow CME ($\sim$320 km s$^{-1}$), observed by the Large Angle and Spectrometric Coronagraph 
(LASCO-C2/C3) onboard the Solar and Heliospheric Observatory (SoHO), with onset at $\sim$05:14 UT, according to linear 
backward-extrapolation of the heights, provided by the LASCO CME Catalog. In addition, an M1.0 SXR flare, observed by the 
Geostationary Operational Environmental Satellites (GOES), with onset at $\sim$05:30 UT and peak at $\sim$05:39 UT, 
was also associated with event 1. Both the CME and the flare associated with event 1 originated from the active region NOAA 
11079, close to the solar limb (S25W84). 

\begin{figure}[!h]
 \centering
 \includegraphics[width=8cm]{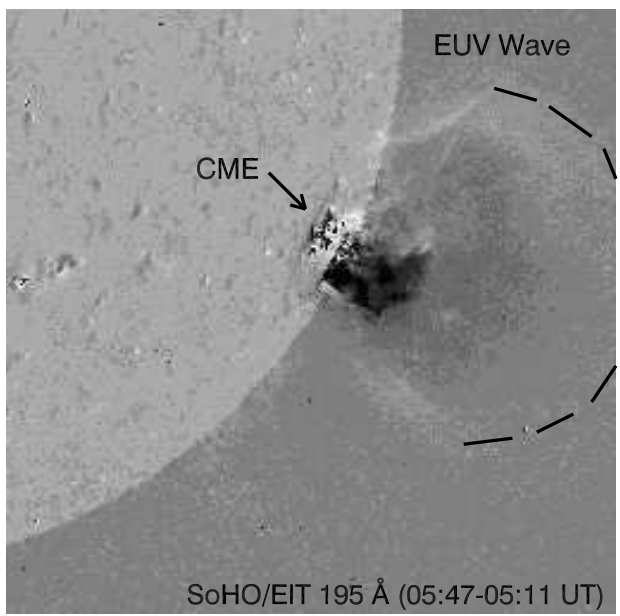}
 \caption{Base-difference image of SoHO/EIT 195 \AA \ showing the EUV wave associated with the type II emission.}
 \label{f2}
\end{figure}

\begin{figure}[!h]
 \centering
 \includegraphics[width=6cm]{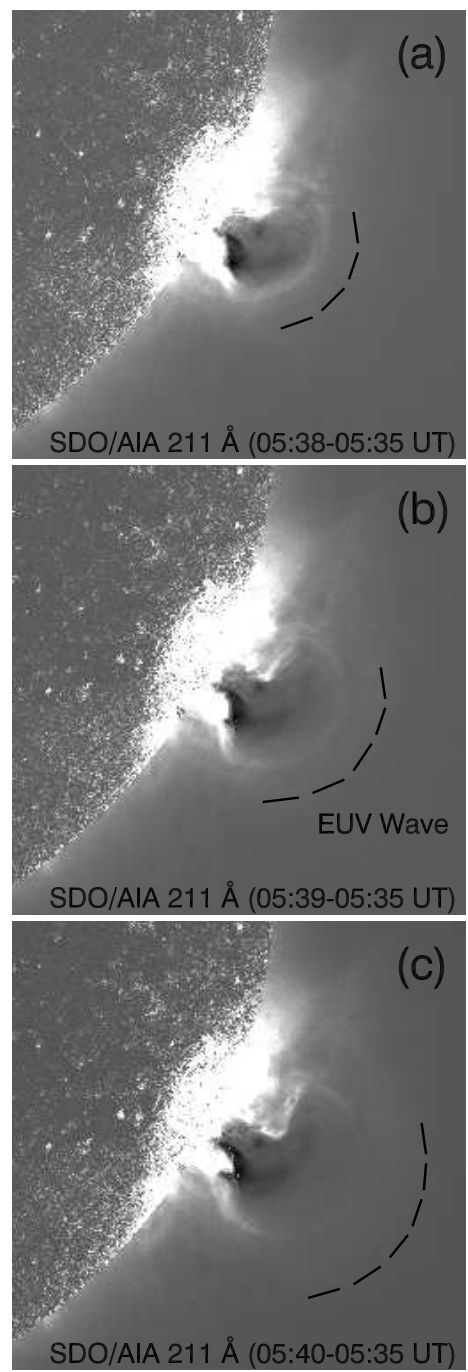}
 \caption{Base-difference images of SDO/AIA 211 \AA \ showing the CME expansion and the EUV wave associated with event 1 
 for: (a) 05:38-05:35 UT; (b) 05:39-05:35 UT; and (c) 05:40-05:35 UT.}
 \label{f3}
\end{figure}

The EUV wave associated with event 1, observed by the Extreme Ultraviolet Imaging Telescope (EIT) onboard SoHO and by 
the Atmospheric Imaging Assembly (AIA) onboard the Solar Dynamics Observatory (SDO) are shown in Figures \ref{f2} and \ref{f3}.

Exponential fits on the intensity maxima of the upper and lower frequency branches of the harmonic lane of event 1 used to estimate 
the frequency drift rate [$\mathrm{d}f/\mathrm{d}t$] of the emission were more consistent with the morphology of the spectrum of the 
event and provided -(1.11\,--\,0.37) MHz s$^{-1}$. This result implies a drift of -(0.55\,--\,0.18) MHz s$^{-1}$ for the fundamental 
emission, taking into account the H/F frequency ratio $f_{\rm{H}}/f_{\rm{F}}=2$. 

The instantaneous band split, given by $BD_{\rm{i}}=f_{\rm{UB}}-f_{\rm{LB}}$, taken at each time step of 0.25 s, was in 
the range of 13\,--\,29 MHz for the harmonic emission of event 1.

\subsection{Event on 6 June 2012} 

\begin{figure}[!h]
 \centering
 \includegraphics[width=9cm]{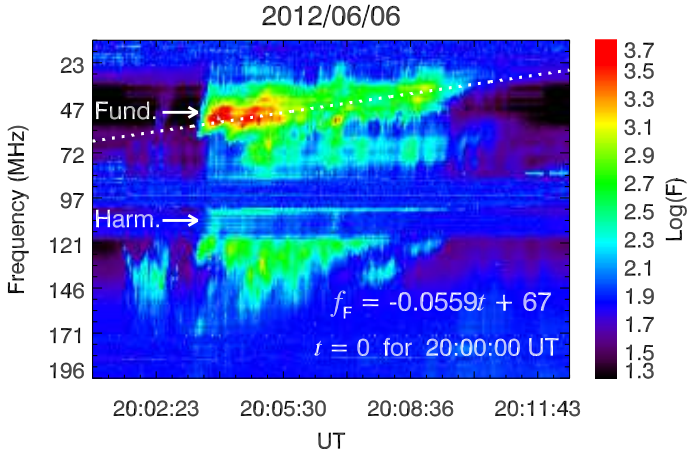}
 \caption{Dynamic spectrum of the type II burst observed by CALLISTO-BIR on June 6, 2012 at $\sim$20:03:25 UT, 
 with linear fit for the fundamental lane.}
 \label{f4}
\end{figure}

The event on 6 June 2012 (hereafter event 2) was observed at $\sim$20:03:25 UT by CALLISTO-BIR (Birr, Ireland) in the operational 
frequency range of 10\,--\,196 MHz. Figure \ref{f4} shows the dynamic spectrum of event 2 with both the fundamental and the harmonic 
emissions, despite the radio frequency interference at $\sim$90\,--\,120 MHz. A type II precursor \citep[see, e.g.,][]{Klassen:1999,Vrsnak:2000} 
at $\sim$20:01:40 UT is also present in the spectrum of event 2. 

Event 2 was associated with a CME with a linear speed of $\sim$494 km s$^{-1}$, observed by LASCO-C2/C3 onboard SoHO, with 
onset at $\sim$19:37 UT, according to linear backward-extrapolation of the heights, provided by the LASCO CME Catalog. An M2.1 
SXR flare, observed by GOES, with onset at $\sim$19:54 UT and peak at $\sim$20:06 UT, was also associated with event 2. Both 
the CME and the flare associated with event 2 originated from the active region NOAA 11494 on the solar disk (S17W08).

\begin{figure}[!h]
 \centering
 \includegraphics[width=6cm]{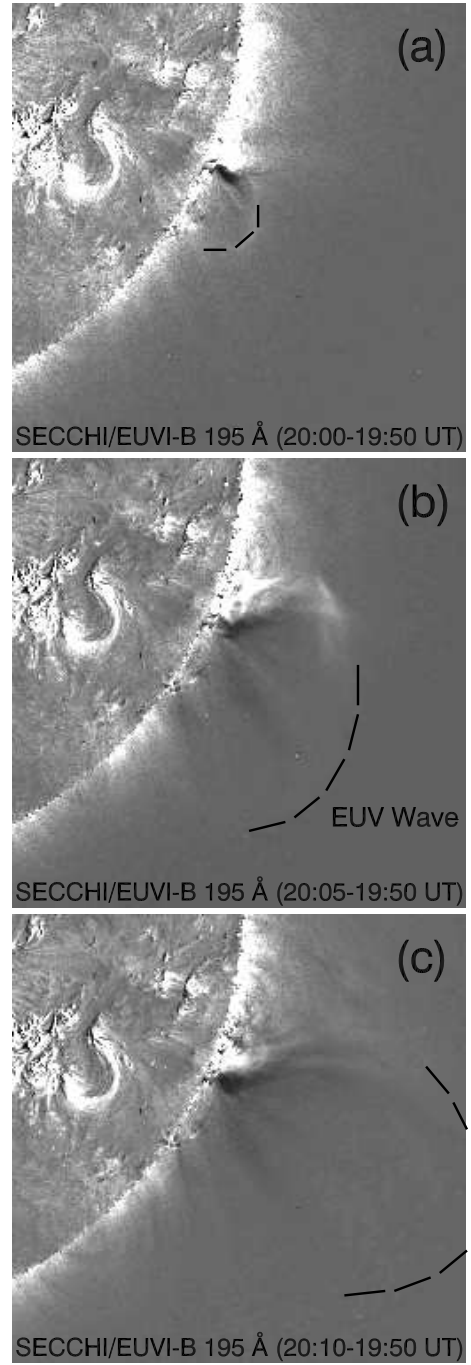}
 \caption{Base-difference images of STEREO/SECCHI/EUVI-B 195 \AA \ for: (b) 20:00-19:50 UT 
 (initial expansion of the CME, $\sim$3 minutes before the onset of the type II emission); (c) 20:05-19:50 UT 
 (fast expansion of the CME during the intensification of the type II emission); and (d) 20:10-19:50 UT (disruption 
 of the CME flux rope at the end of the type II emission).}
 \label{f5}
\end{figure}

The expansion of the CME associated with event 2 was clear from the EUV images provided by the Extreme Ultraviolet Imager (EUVI) 
onboard the Solar Terrestrial Relations Observatory (STEREO) presented in Figure \ref{f5}. 

Event 2 showed an intense fundamental emission with a herringbone-like feature \citep[see, e.g.,][]{Mann:1995a} and with 
a linear-like frequency drift. A linear fit on the intensity maxima of the fundamental lane provided a frequency drift rate of -0.056 MHz s$^{-1}$. 
Event 2 presented an instantaneous bandwidth [$\Delta f_{\rm{i}}$] of 5\,--\,26 MHz for the fundamental lane, adopting the 
half-flux bandwidths of the maxima of the emission. Since there was no band-splitting for event 2, a different approach 
was used, and the instantaneous bandwidth was adopted in place of the instantaneous band split, similar to that used by \citet{Mann:1995b}.

\section{Physical parameters: models and equations} 

From the observational parameters of the type II bursts, it is possible, with some assumptions, to estimate the physical 
parameters of the radio sources. In this section, we present the models and equations adopted in this article.

In most cases, the use of the 1\,--\,2$\times$\citet{Newkirk:1961} density model to estimate the radio source speed is 
compatible with a radial shock motion \citep[see, e.g.,][]{Vrsnak:2001}, whereas the use of the 3\,--\,4$\times$\citet{Newkirk:1961} 
model is suitable for an oblique propagation of the type-II-emitting shock segment \citep[see, e.g.,][]{Klassen:1999,Classen:2002,Cunha-Silva:2014}, 
avoiding the underestimation of the shock speed. The \citet{Newkirk:1961} density model is given by   

\begin{equation} 
n_{\rm{e}}=n_{\rm{e}_{0}}\times10^{4.32\left(\frac{\rm{R_{\odot}}}{R}\right)}, \label{eq1}
\end{equation}
where $n_{\rm{e}_{0}}=4.2\times10^{4}$ cm$^{-3}$, $R$ is the heliocentric distance and $\rm{R_{\odot}}$ is the solar radius. 

\citet{Cairns:2009} extracted $n_{\rm{e}}(R)$ directly from coronal type III radio bursts for $40\leq f\leq180$ MHz and found 
$n_{\rm{e}}\propto(R-\rm{R_{\odot}})^{-2}$ for $R<2\rm{R_{\odot}}$ (note that for $R\gg R_{\odot}$ this falloff becomes 
$n_{\rm{e}}\propto R^{-2}$, which is a standard solar wind property that holds for $R\geq10\rm{R_{\odot}}$). 
Thus, we adopted

\begin{equation} 
n_{\rm{e}}(R)=C(R-\rm{R_{\odot}})^{-2}, \label{eq2}
\end{equation}
where $C$ is a constant determined from the time-varying radiation frequency and the radial speed of the source.

We compared these two different density-height models applied to the type II events with the heights of the 
associated EUV waves with the aim of determining which model is more appropriate to be adopted in estimating the physical 
parameters of the radio sources.

A practical expression for the radial speed of the shock [$\mathrm{d}R/\mathrm{d}t$], for $n\times$\citet{Newkirk:1961} 
density model, is \citep{Cunha-Silva:2014}

\begin{equation} 
\frac{\mathrm{d}R}{\mathrm{d}t}= \frac{-6.04\times10^{6}\frac{\mathrm{d}f}{\mathrm{d}t}}{\mathrm{ln} \ 10\times\Bigl(2\mathrm{log}(f_{\rm{p}}) - 
\mathrm{log}(3.39n)\Bigr)^{2}f_{\rm{p}}}. \label{eq3}
\end{equation}

The density jump at the shock front [$X_{\rm{n}}$] is determined from the relative instantaneous band split 
[$BD_{\rm{i}}/f_{\rm{LB}}$] \citep{Vrsnak:2002}:

\begin{equation} 
X_{\rm{n}}=\left(\frac{BD_{\rm{i}}}{f_{\rm{LB}}}+1\right)^{2}=\left(\frac{f_{\rm{UB}}}{f_{\rm{LB}}}\right)^{2}. \label{eq4}
\end{equation}  

For a shock perpendicular to the magnetic field and $\beta\rightarrow0$, the Alfv\'en Mach number [$M_{\rm{A}}$] is \citep{Vrsnak:2002}

\begin{equation} 
M_{\rm{A}}=\sqrt{\frac{X_{\rm{n}}(X_{\rm{n}}+5)}{2(4-X_{\rm{n}})}}. \label{eq5}
\end{equation}   

The Alfv\'en speed [$v_{\rm{A}}$], in turn, can be obtained directly from the shock speed [$v_{\rm{s}}$] and $M_{\rm{A}}$:

\begin{equation} 
v_{\rm{A}}=\frac{v_{\rm{s}}}{M_{\rm{A}}}. \label{eq6}
\end{equation}

Finally, the magnetic-field strength [$B$] in regions of shock surrounding a CME can be estimated using \citep{Kim:2012} 

\begin{equation} 
B=5.0\times10^{-7}v_{\rm{A}}\sqrt{n_{\rm{e}}} \ \ [\mathrm{G}]. \label{eq7}
\end{equation} 

\section{Results and discussion} 

We compared the heights of the EUV wave fronts with the heights of the radio sources and then estimated the physical parameters of the radio 
sources by applying the more appropriate coronal density model, taking into account that under the assumed band-split interpretation for event 1, 
the application of coronal density models to the lower-frequency branch is more appropriate to estimate the shock kinematics, since it represents 
the emission from undisturbed corona. 

\subsection{Event 1} 

\begin{figure}[!h]
 \centering
 \includegraphics[width=9.2cm]{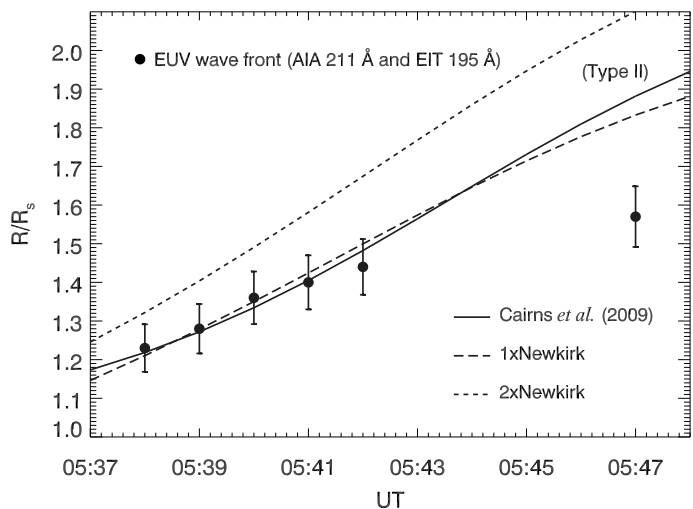}
 \caption{Comparison of the type II heights, obtained with the density models of 1\,--\,2$\times$Newkirk (dashed lines) and Cairns et al. (2009) 
 (solid line), applied to the frequencies provided by the exponential fit on the intensity maxima of the lower-frequency branch, with the heights 
 of the EUV wave fronts (five data points from SDO/AIA 211 \AA, limited by the field of view (FOV) of the instrument, and one data point 
 from SoHO/EIT 195 \AA, at the end of the type II emission), for event 1. The models of 1$\times$Newkirk and of Cairns et al. (2009) provided 
 similar results for event 1.}
 \label{f6}
\end{figure}

Figure \ref{f6} shows the type II heights, obtained with the density models of 1\,--\,2$\times$\citet{Newkirk:1961} and of \citet{Cairns:2009}, 
and the heights of the EUV wave fronts associated with event 1. We adopted 10\% of the measures for the error bars of the EUV wave fronts, considering 
the uncertainties of 7\,--\,12\% in the measurements.

Figure \ref{f6} shows that the heights of the EUV wave fronts associated with event 1 are consistent with the heights of the type II 
emission obtained with the density models of 1$\times$\citet{Newkirk:1961} and of \citet{Cairns:2009} for the first five minutes of the event, 
supporting a quasi-radial propagation of the type-II-emitting shock segment. 

\begin{figure}[!h]
 \centering
 \includegraphics[width=9.0cm]{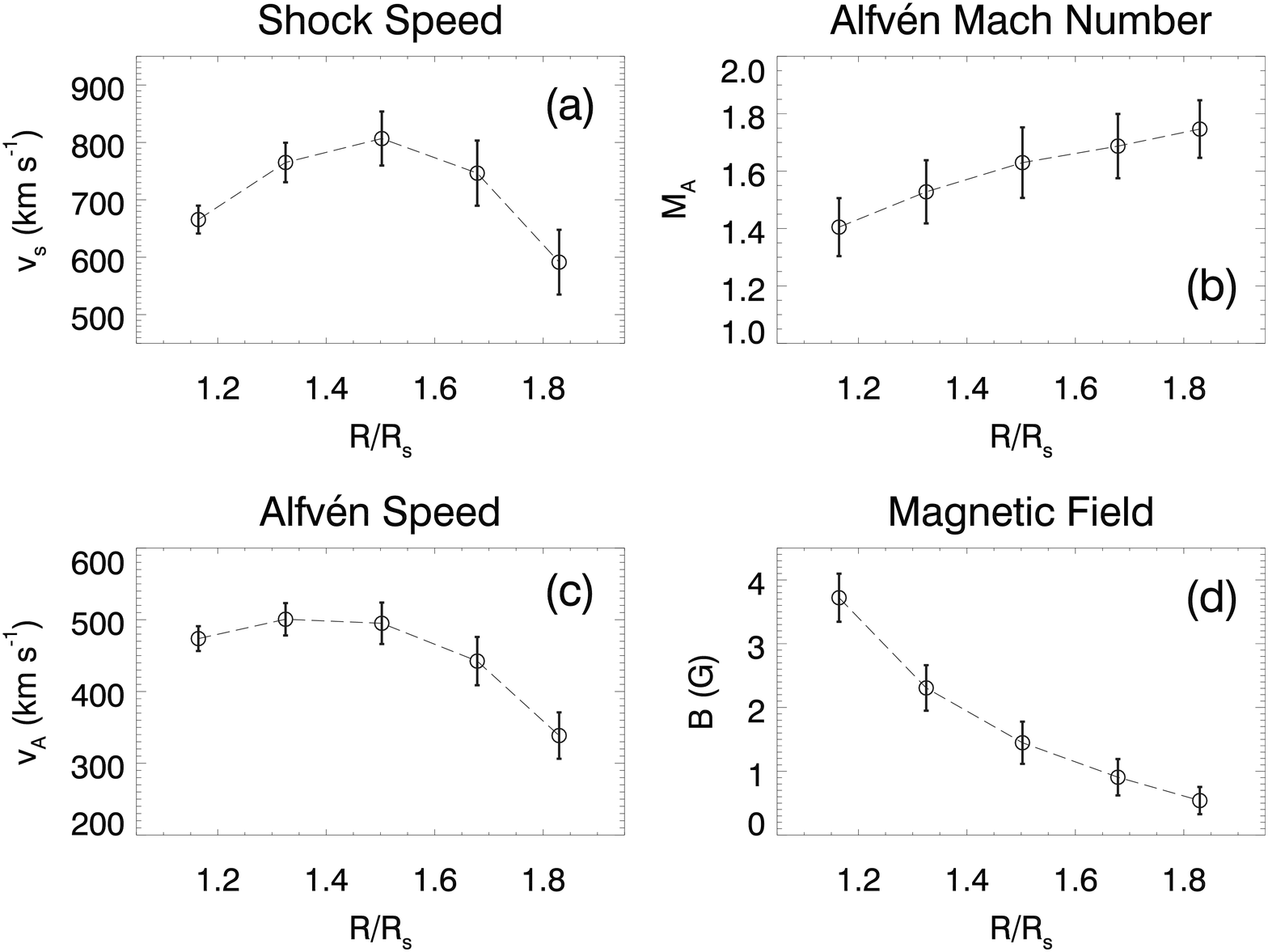}
 \caption{Physical parameters of the radio source for event 1: (a) shock speed, calculated from the frequency drift [$\mathrm{d}f/\mathrm{d}t$] 
 and the plasma frequency [$f_{\rm{p}}$]; (b) Alfv\'en Mach number, calculated from the density jump at the shock front [$X_{\rm{n}}$]; 
 (c) Alfv\'en speed, calculated from the shock speed [$v_{\rm{s}}$] and the Alfv\'en Mach number [$M_{\rm{A}}$]; and (d) magnetic-field 
 strength, calculated from the electron number density [$n_{\rm{e}}$] and the Alfv\'en speed [$v_{\rm{A}}$]. The values were determined 
 with respect to the central frequencies of the lower-frequency branch of the radio emission.}
 \label{f7}
\end{figure}

Given that the models of 1$\times$\citet{Newkirk:1961} and of \citet{Cairns:2009} provided results consistent with the heights of the 
EUV wave fronts for event 1, we adopted the former to estimate the physical parameters of the radio source for this event. In Figure \ref{f7}, 
the physical parameters of the radio source for event 1 are presented as a function of the heliocentric distance. The error bars of 
the estimates were defined as the instrumental errors of the measures, considering a quadratic error equal to zero.

Figure \ref{f7}(a) shows that a shock speed of 590\,--\,810 km s$^{-1}$ as found for event 1 is consistent with the initial and average speeds of 
the EUV wave fronts of 720 and 610 km s$^{-1}$. Our results show an accelerating shock with a speed of 665\,--\,810 km s$^{-1}$ during the 
first five minutes of the type II emission, followed by a shock deceleration, with speeds of 810 to 590 km s$^{-1}$ for the following five 
minutes of the event, which is consistent with the strong acceleration (up to speeds of $\sim$400 km s$^{-1}$) followed by the deceleration 
of the EUV bubble associated with the CME, found by \citet{Patsourakos:2010}. A similar behavior was reported by \citet{Gopalswamy:2012}, who 
found an accelerating shock with speeds of $\sim$590\,--\,720 km s$^{-1}$ for the first three minutes of the type II emission, followed by a 
shock deceleration with speeds of $\sim$720 to 640 km s$^{-1}$ for the next two minutes. The adoption of the central frequencies of the 
lower-frequency branch to estimate the heights of the radio source has accounted for our higher values. However, our results are different 
from those obtained by \citet{Vasanth:2014}, who found a sharp decrease in the shock speed, from 820 km s$^{-1}$ to 580 km s$^{-1}$, for 
05:39\,--\,05:41 UT (three and five minutes after the onset of the type II emission). Our estimates of the frequency drift rate using exponential 
fits on the intensity maxima of the radio emission may be responsible for these differences. 

A slightly increasing Alfv\'en speed of 470\,--\,500 km s$^{-1}$ for the first three minutes of event 1, followed by a decrease from 
500 to 340 km s$^{-1}$, shown in Figure \ref{f7}(c), was the result of a direct determination of this parameter from the shock speed 
and the Alfv\'en Mach number, and is marginally compatible with the typical decrease of this parameter in regions of shock formation 
($R<2\rm{R_{\odot}}$) \citep[see, e.g.,][]{Vrsnak:2002}. This finding, however, is somewhat different from that reported by \citet{Gopalswamy:2012}, 
who found a sharp increase in the Alfv\'en speed of $\sim$140\,--\,460 km s$^{-1}$ for the first three minutes of the type II emission, followed by 
a decrease from $\sim$460 to 410 km s$^{-1}$. The values found by us and by \citet{Gopalswamy:2012} are below those reported by \citet{Vrsnak:2002}, 
who found 600\,--\,1000 km s$^{-1}$ for the heliocentric distances of $\sim$1.2\,--\,1.8, applying a fifth degree polynomial fit on measurements 
based on observational data.     

A magnetic field strength of 0.5\,--\,3.7 G, for the heliocentric distances of $\sim$1.2\,--\,1.8, found for event 1, is consistent with the 
values of about 1.2\,--\,5.0 G reported by \citet{Vrsnak:2002} for these heliocentric distances, obtained by a power-law fit on measurements 
based on observational data, assuming the coronal density model of 2$\times$\citet{Newkirk:1961}. Considering the error bars of our estimates, 
shown in Figure \ref{f7}(d), a magnetic field strength of 2.3 G for a shock speed of $\sim$760 km s$^{-1}$, obtained for three minutes after 
the onset of the type II emission, is consistent with the magnetic field strength of 1.7\,--\,1.9 G found by \citet{Kouloumvakos:2014} for 
a shock speed of $\sim$700 km s$^{-1}$. Our higher values for the initial heights of the type II emission are due to our higher values for 
the Alfv\'en speed, even with our higher Alfv\'en Mach number of 1.4\,--\,1.7, against 1.3\,--\,1.5, found by \citet{Kouloumvakos:2014}. 
Our values for the Alfv\'en Mach number are consistent with the measurements of about 1.1\,--\,1.9 reported by \citet{Vrsnak:2002} for 
the heliocentric distances of $\sim$1.3\,--\,2.5.

\subsection{Event 2} 

\begin{figure}[!h]
 \centering
 \includegraphics[width=9.2cm]{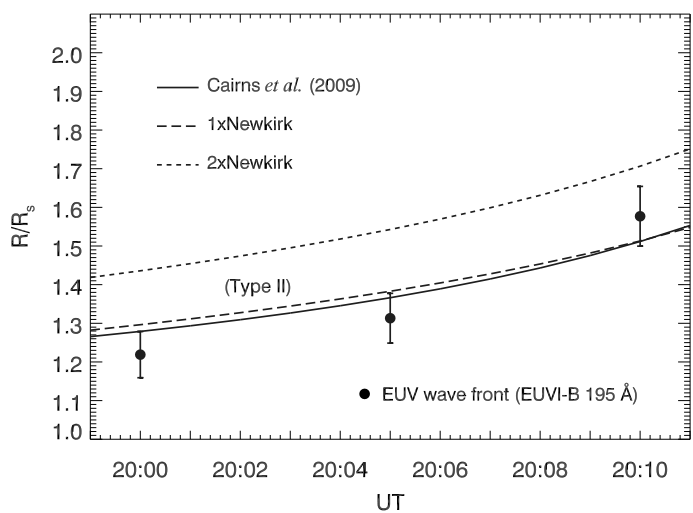}
 \caption{Comparison of the type II heights obtained with the density models of 1\,--\,2$\times$Newkirk (dashed lines) and of Cairns et al. (2009) 
 (solid line), applied to the frequencies provided by the linear fit on the intensity maxima of the fundamental emission, with the heights of the 
 EUV wave fronts (three data points from STEREO/EUVI-B 195 \AA) for event 2. Similarly as for event 1, the models of 1$\times$Newkirk and of 
 Cairns et al. (2009) are equivalent for event 2.}
 \label{f8}
\end{figure}

Taking the analysis of event 1 as a touchstone for the analysis of event 2, we compared the heights of the EUV wave fronts with the heights of the 
radio source and determined the physical parameters of the radio source with the more appropriate model, adopting similar criteria and the same 
equations. Figure \ref{f8} shows the type II heights, obtained with the density models of 1\,--\,2$\times$\citet{Newkirk:1961} and of \citet{Cairns:2009} 
and the heights of the EUV wave fronts associated with event 2.

As found for event 1, the heights of the EUV wave fronts associated with event 2 are consistent with the heights of the type II emission 
obtained with the density models of 1$\times$\citet{Newkirk:1961} and of \citet{Cairns:2009}, as is clear from Figure \ref{f8}, supporting 
a quasi-radial propagation of the type-II-emitting shock segment. In this case, however, we found a radio source moving away from the Sun 
more slowly than the EUV wave fronts. 

\begin{figure}[!h]
 \centering
 \includegraphics[width=9.0cm]{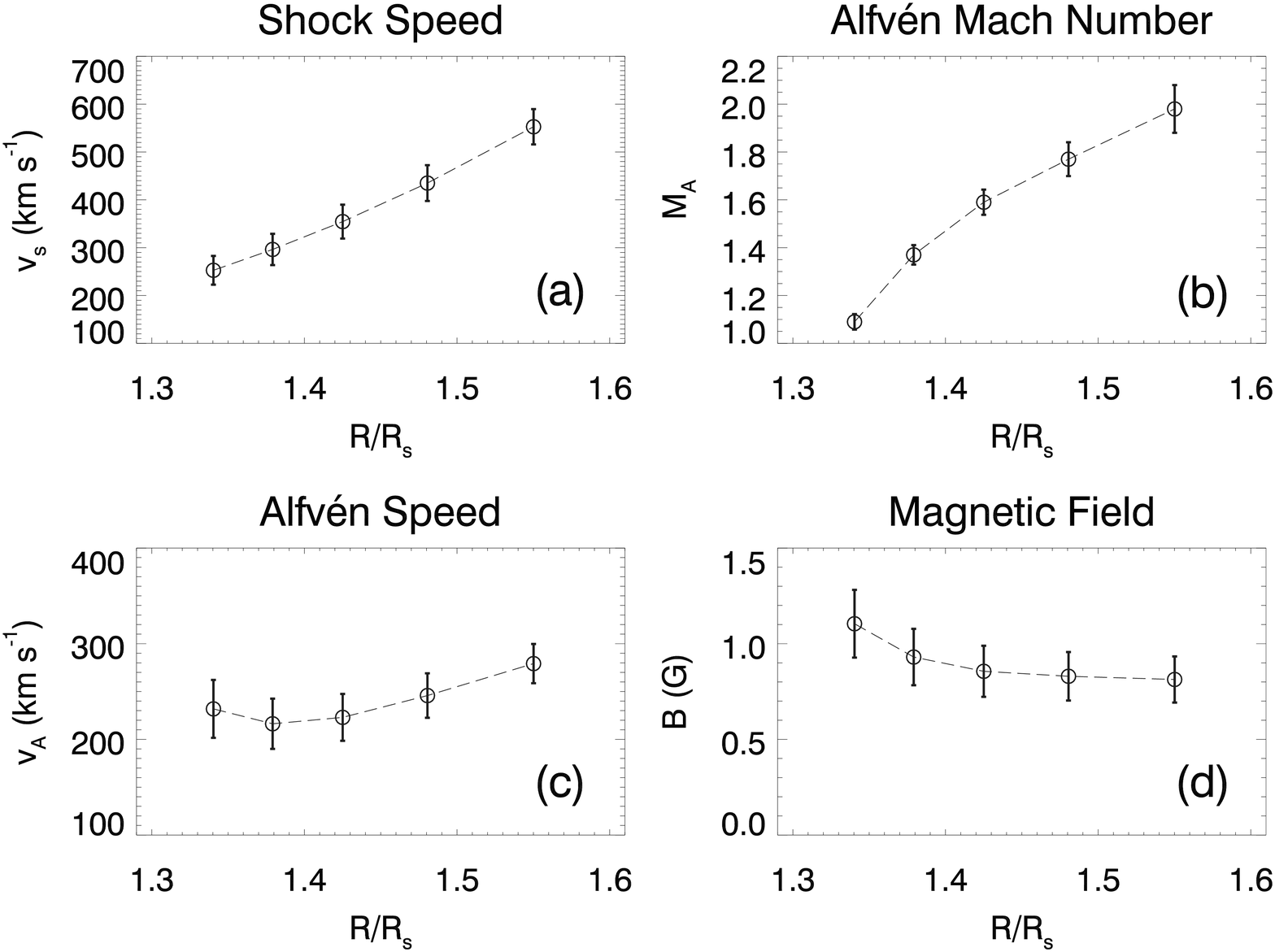}
 \caption{Physical parameters of the radio source for event 2: (a) shock speed, calculated from the frequency drift [$\mathrm{d}f/\mathrm{d}t$] 
 and the plasma frequency [$f_{\rm{p}}$]; (b) Alfv\'en Mach number, calculated from the density jump at the shock front [$X_{\rm{n}}$]; 
 (c) Alfv\'en speed, calculated from the shock speed [$v_{\rm{s}}$] and the Alfv\'en Mach number [$M_{\rm{A}}$]; and (d) magnetic-field 
 strength, calculated from the electron number density [$n_{\rm{e}}$] and the Alfv\'en speed [$v_{\rm{A}}$]. The values were determined 
 with respect to the lower frequencies of the fundamental lane of the radio emission.}
 \label{f9}
\end{figure}

The morphology of the spectrum of event 2, with a linear-like frequency drift, accounts for the greater similarity between the results for the heights 
of the radio source provided by the density models of 1$\times$\citet{Newkirk:1961} and of \citet{Cairns:2009}. Again, we adopted the former to estimate 
the physical parameters of the radio source of event 2. In Figure \ref{f9}, the physical parameters of the radio source for event 2 are 
presented as a function of the heliocentric distance.

In Figure \ref{f9}(a), an accelerating shock with speeds of 250\,--\,550 km s$^{-1}$ indicates radio emission from the early stages of a shock 
driven by CME expansion during its fast acceleration phase. This interpretation is consistent with the linear speed of the associated CME of 
490 km s$^{-1}$, which is higher than the average speed of the associated EUV wave fronts of 420 km s$^{-1}$. 

The Alfv\'en speed determined directly from the shock speed and the Alfv\'en Mach number, according to Eq. \ref{eq6}, showed a somewhat 
atypical behavior for event 2 that is due to a combination of an increasing shock speed with an increasing Alfv\'en Mach number of 1.1\,--\,2.1. 
Different from what was obtained for event 1, we found a slightly decreasing Alfv\'en speed of 230 to 220 km s$^{-1}$, followed by an 
increase of 220\,--\,280 km s$^{-1}$, shown in Figure \ref{f9}(c). Our values for the Alfv\'en Mach number are consistent with the 
measurements of about 1.1\,--\,1.9 reported by \citet{Vrsnak:2002} for heliocentric distances of $\sim$1.3\,--\,2.5.

Another interesting feature of Figure \ref{f9}(d), is a comparatively slow decrease of the magnetic field strength, which becomes almost 
constant between $R=1.45$ and $1.55$. This somewhat atypical behavior could be attributed to the fact that the magnetic field strength 
estimate depends not only on the coronal density, but also on the estimate of the Alfvén speed based on Eq. \ref{eq7}, which accounts 
for this somewhat atypical behavior.

\section{Conclusions}
  
We investigated two solar type II radio bursts associated with EUV waves, one of which included a split-band event, with the fundamental emission 
partially reabsorbed, that took place close to the limb, and one event without band-splitting, with both the fundamental and harmonic emissions 
intense on the spectrum, that took place on the disk. Both events agreed with the observational pattern reported by \citet{Biesecker:2002} that 
type II events on the disk are often observed in the fundamental and harmonic emissions, whereas only the harmonic emission is usually observed 
for those at the limb.      

For both events, the heights of the EUV wave fronts \textit{are} compatible with the heights of the radio source obtained with the 
density models of 1$\times$\citet{Newkirk:1961} and of \citet{Cairns:2009}, supporting a quasi-radial propagation of the type-II-emitting 
shock segment for the events.

The finding of an accelerating shock during the first five minutes of event 1, followed by a shock deceleration for the next five 
minutes of the event, is consistent with the strong acceleration followed by deceleration of the EUV bubble associated with the event 
found by \citet{Patsourakos:2010} and is also consistent with a similar shock speed behavior reported by \citet{Gopalswamy:2012} for 
the first five minutes of the event. 

The finding of an accelerating shock throughout event 2 proved to be consistent with the kinematics of the EUV wave fronts, which we 
found to be moving faster than the radio source.

Our results support a close association between the radio source and the CME expansion for both events, as shown by the close relation 
between the type-II-emitting shock segment and the EUV wave, which clearly was produced by the CME.

\begin{acknowledgements}
      The authors acknowledge financial support from the S\~ao Paulo Research Foundation (FAPESP), grant number 2012/08445-9. CLS acknowledges 
      financial support from the S\~ao Paulo Research Foundation (FAPESP), grant number 2014/10489-0. RDCS acknowledges a scholarship from the 
      S\~ao Paulo Research Foundation (FAPESP), grant number 2012/00009-5, and financial support from UNIVAP-FVE. FCRF thanks CNPq for the 
      scholarship granted under process 308755/2012-0. The authors are grateful to the e-CALLISTO science teams for the solar data. The EUV 
      images were courtesy of ESA/NASA/ SoHO/EIT, NASA/STEREO/EUVI, and NASA/SDO/AIA consortia. The authors are grateful to the 
      anonymous referee for the significant and constructive remarks that helped to improve the quality and presentation of this article. 

\end{acknowledgements}

%-------------------------------------------------------------------

\bibliographystyle{aa} %% style aa.bst
% \bibliography{/home/rafael/cunha_silva_aea1_astroph/references}  

\end{document}